\documentclass[12pt,preprint]{aastex}

\newcommand{\kTe}{$kT_{\rm e}$}

\shorttitle{{\it Chandra} Observations of the Cygnus Loop NE Rim}
\shortauthors{Katsuda et al.}

\begin{document}


\title{{\it Chandra} Observations of the Northeastern Rim of the
  Cygnus Loop}


\author{Satoru Katsuda\altaffilmark{1}, Hiroshi Tsunemi\altaffilmark{1},
Masashi Kimura\altaffilmark{1}, and Koji Mori\altaffilmark{2}}
\email{katsuda@ess.sci.osaka-u.ac.jp, tsunemi@ess.sci.osaka-u.ac.jp,
mkimura@ess.sci.osaka-u.ac.jp, mori@astro.miyazaki-u.ac.jp} 


\altaffiltext{1}{Department of Earth and Space Science, Graduate
  School of Science, Osaka University, 1-1 Machikaneyama, Toyonaka,
  Osaka 560-0043, Japan}
\altaffiltext{2}{Department of Applied Physics, Faculty of Engineering,
University of Miyazaki, 1-1 Gakuen Kibana-dai Nishi, Miyazaki, 889-2192,
Japan}


\begin{abstract}

 We present results from spatially resolved spectral analyses of the
 northeastern (NE) rim of the Cygnus Loop supernova remnant (SNR)
 based on two {\it 
 Chandra} observations.  One pointing includes northern outermost
 abundance-enhanced regions discovered by recent {\it
 Suzaku} observations, while the other pointing is located on regions
 with ``normal'' abundances in the NE rim of the Cygnus Loop.  The
 superior spatial resolving power of {\it Chandra} allows us to reveal
 that the abundance-enhanced region is concentrated in a
 $\sim$200$^{\prime\prime}$-thickness region behind the shock front. 
 We confirm absolute metal abundances (i.e., relative to H) as well
 as abundance ratios between metals are consistent with those of the
 solar values within a factor of $\sim$2.  Also, we find that the emission
 measure in the region gradually decreases toward the shock front.
 These features are in contrast with those of the ejecta fragments around
 the Vela SNR, which leads us to believe that the abundance
 enhancements are not likely due to metal-rich ejecta.  We suggest
 that the origin of the plasma in this region is the interstellar
 medium (ISM).  In the ``normal''  
 abundance regions, we confirm that abundances are depleted to the
 solar values by a factor of $\sim$5 that is not expected in the ISM
 around the Cygnus Loop.  Introduction of non-thermal emission in our
 model fitting can not naturally resolve the abundance-depletion problem.
 The origin of the depletion still remains as an open question.  

\end{abstract}


\keywords{ISM: abundances --- ISM: individual (Cygnus Loop) --- supernova
remnants --- X-rays: ISM }

\section{Introduction}

The Cygnus Loop is one of the nearest (540 pc: Blair et al.\ 2005)
supernova remnants (SNRs). The angular dimensions are
$2^\circ.8\times3^\circ.5$ (Leahy et al.\ 1997) and the age is
considered to be $\sim$10000 yrs.  The low column density of the
foreground material makes it an ideal target to study UV and soft
X-ray emission from the remnant. 
The X-ray boundary in the northeastern (NE) rim of the Cygnus Loop is
associated with Balmer-dominated filaments which mark current
locations of the blast wave (Chevalier 1978).  Balmer-dominated
filaments in this region have been well studied by optical as well as UV
observations (e.g., Raymond et al.\ 1983; Sankrit et al.\ 2000;
Ghavamian et al.\ 2001).  Hester et al.\ (1994) studied a 
Balmer-dominated filament in detail and reported that it was recently
(in past 1000 yrs) decelerated from $\sim$400 km\,sec$^{-1}$ to
$\sim$180 km\,sec$^{-1}$.  The rapid deceleration of the shock
velocity was considered to be a result of the blast wave hitting the
wall of a cavity which 
surrounded the supernova precursor.  The cavity wall/blast wave
interaction was also suggested by X-ray observations.  For example,
Levenson et al.\ (1999) revealed that a soft spatially thin ($<5'$) shell
surrounded almost the entire rim of the Loop based on a {\it ROSAT} PSPC
hardness map.   They concluded that the soft shell occurred where the
cavity wall decelerated the blast wave.

{\it ASCA} observations of the NE rim of the
Cygnus Loop revealed low abundances relative to the solar values by a
factor of $\sim$5 (Miyata \& Tsunemi 1999).  A follow-up {\it Suzaku} 
observation of the region confirmed the depleted abundances there
(Miyata et al.\ 2007).  {\it Chandra} observation of the southwestern
rim of the Cygnus Loop revealed that oxygen abundance there is also
depleted by the same factor as that observed in the NE rim (Leahy
2004).  The low abundances were considered to be a common feature in
the rim of the Cygnus Loop.  However, recent {\it Suzaku} observations 
of the northern outermost region in NE rim of the Cygnus Loop revealed
enhanced abundances: a white dashed polygon in Fig.~\ref{fig:HRI
image} showed high metal abundances relative to those in the other
region: C$\sim$0.7, N$\sim$0.7, O$\sim$0.4, Ne$\sim$0.6, Mg$\sim$0.3,
and Fe$\sim$0.3 times the solar values (Katsuda et al.\ 2008).  
Neither a circumstellar medium, fragments of ejecta, nor abundance
inhomogeneities of the local interstellar medium around the Cygnus
Loop can explain the relatively enhanced abundance in the region. 
The origin of the abundance inhomogeneity in the NE rim remained as an 
open question.   

Here, we present results from {\it Chandra} observations of the NE rim
of the Cygnus Loop.  Utilizing the high spatial resolving power
of {\it Chandra}, we reveal detailed spatial distributions of metal
abundances, emission measure, and thermodynamic parameters from our
spatially resolved spectral analysis.  In this paper, we attempt to
reveal the origin of the abundance enhancements in the northern
outermost region of the NE rim as well as to understand the cause of
the abundance depletion in the rest of the NE rim. 

\section{Observations}

\begin{figure}
\includegraphics[angle=0,scale=.6]{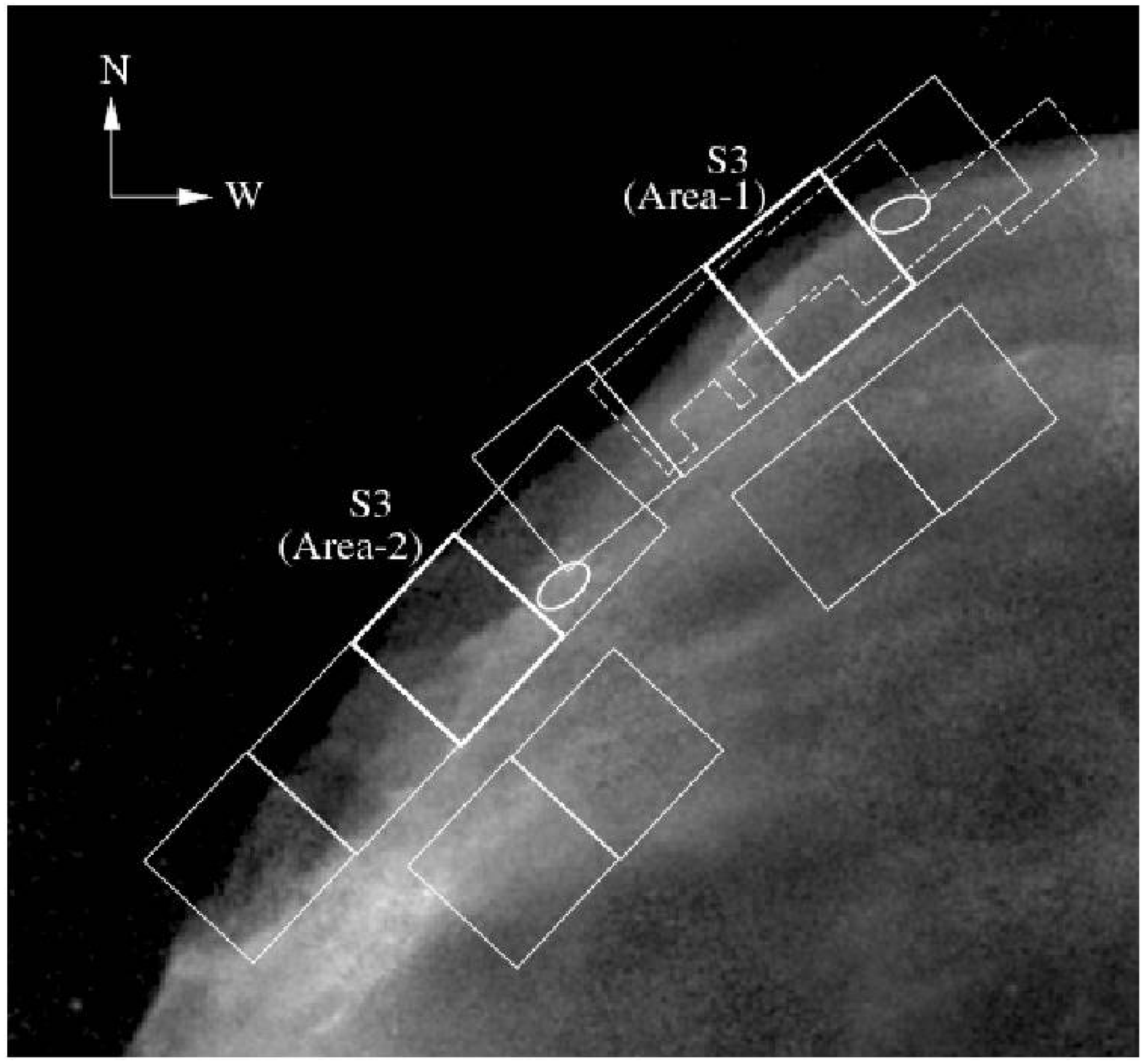}
\caption{Exposure-corrected {\it ROSAT} HRI image of the NE quadrant
 of the Cygnus Loop.  The FOV of {\it Chandra} are overlaid on this
 image as white solid boxes.  A white dashed polygon shows the
 abundance-enhanced region discovered by {\it Suzaku} (Katsuda et al.\
 2008).  Two ellispes are the regions where we extract spectra other
 than S3 chips.} 
\label{fig:HRI image}
\end{figure}

{\it Chandra} Advanced CCD Imaging Spectrometer (ACIS) observed the
NE rim of the Cygnus Loop in two pointings (PI: Gaetz, T.).  The
nominal position of one pointing (ObsID.\ 2821), which covers the
northern portion of the 
NE rim, is [(RA,DEC)=(20$^h$54$^m$38$^s$.8,
  32$^\circ$16$^\prime$27$^{\prime\prime}$.9) J2000], while that of
the other pointing (ObsID.\ 2822), which covers the southern portion of
the NE rim, is [(RA,DEC)=(20$^h$56$^m$10$^s$.6,
  31$^\circ$55$^\prime$19$^{\prime\prime}$.3) J2000].  The observation
dates are 2001 December 24 and 16 for ObsID.\ 2821 and 2822,
respectively.  The total fields of view (FOV) of twelve CCDs for the two
pointings are shown as white boxes in Fig.~\ref{fig:HRI image}.
We start our analysis from level 2 event files processed with
calibration data files in CALDB ver.\ 3.3.0.  We exclude the
high-background periods for data from ObsID.\ 2821.  On the other
hand, there seems no significant background flares for the data from
ObsID.\ 2822 so that we reject no data from the level 2 event file for
our analysis.  The resulting net exposure times for ObsID.\ 2821 and
2822 are 35\,ks and 59\,ks, respectively.

\section{Spatially Resolved Spectral Analysis}

\begin{figure}
\includegraphics[angle=0,scale=0.8]{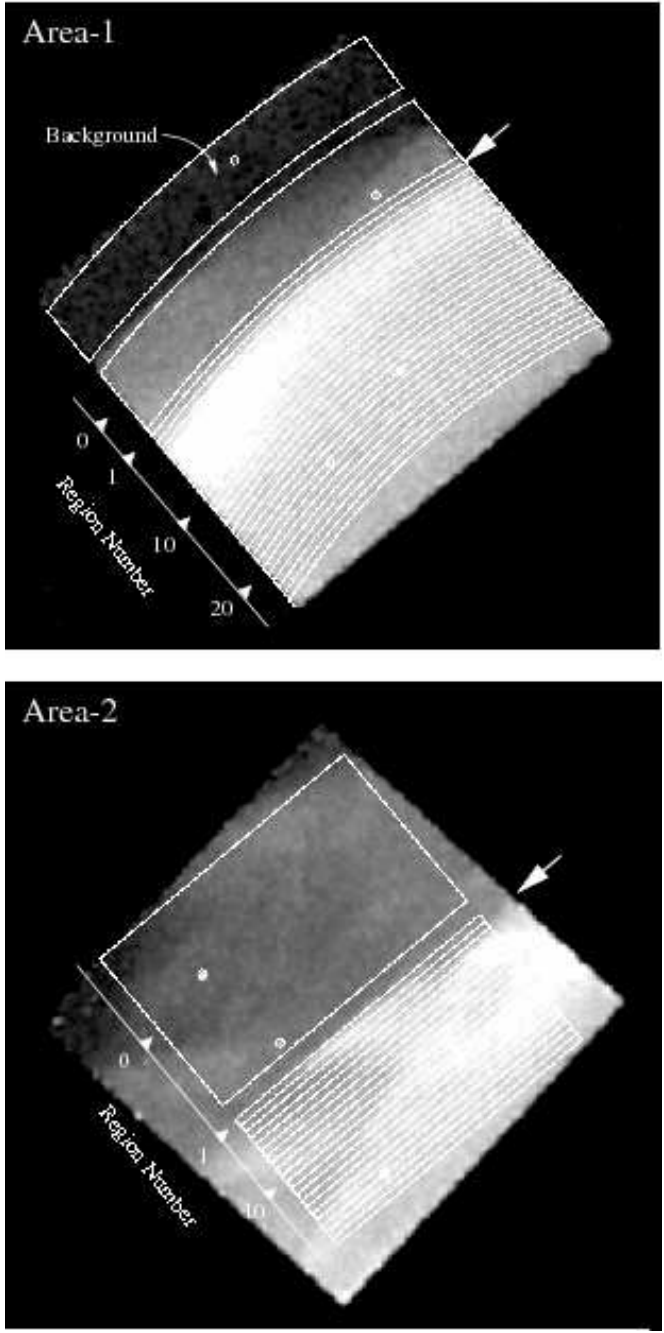}
\caption{{\it Upper}: Exposure and vignetting corrected {\it Chandra}
 S3 chip image for ObsID.\ 2821.  The spectral extraction regions are
 shown as white annuli.  Several point sources indicated by white
 circles are excluded from our spectral analysis.  An arrow seen in the
 upper right indicates the position of the discontinuity of the X-ray
 intensity.  {\it Lower}: Same as Fig.~\ref{fig:Chandra image} {\it upper}
 but for ObsID.\ 2822.}    
\label{fig:Chandra image}
\end{figure}

In order to investigate the post-shock plasma structures in detail, we
perform spatially resolved spectral analysis for the cleaned {\it
  Chandra} data.  To this end, we concentrate on the data taken
  by the S3 (Back-illuminated: BI) chip, since its energy resolving
  power is better than that of another BI (S1) chip, and the numbers
  of photons detected in BI chips are richer than those in
  Front-illuminated (FI) chips for the spectra from the Cygnus Loop.
Figure~\ref{fig:Chandra image} shows ACIS S3 chip images.  
We hereafter name the area covered by the S3 chip in ObsID.\ 2821 
as ``area-1'' (see, Fig.~\ref{fig:Chandra image} {\it upper}),
while that covered by the S3 chip in ObsID.\ 2822 as ``area-2''
(see, Fig.~\ref{fig:Chandra image} {\it lower}).  Area-1
includes the abundance-enhanced region, while area-2 is located on the
region with ``normal'' abundances in the NE rim of the Cygnus Loop.
As indicated by arrows in Fig.~\ref{fig:Chandra image},
discontinuities in X-ray intensity can be seen inside the X-ray
boundaries in both of the two areas.  This fact suggests that the
shock structures in these regions can be considered as multiple shocks
along the line of sight.  We extract spectra from
20$^{\prime\prime}$-spaced (annular for area-1, rectangle for area-2)
regions inside the discontinuities of X-ray intensity (from Reg-1 to
Reg-26/19 for area-1/2).  We arrange the regions such that each 
region overlaps with each other by half of its size.  Outside the
discontinuities, X-ray surface brightness is relatively low so that we
extract spectra from only one region to obtain enough photons to
perform spectral analysis in good statistics.  The region is named as
Reg-0.  We show all the spectral extraction regions in
Fig.~\ref{fig:Chandra image}.   

We subtract background emission from a source free (i.e., outside
the X-ray boundary of the Cygnus Loop) region in the same chip for
area-1 (see, Fig.~\ref{fig:Chandra image} {\it upper}).  On the other
hand, it is quite difficult to subtract background emission from the
same chip for area-2, since the X-ray emission from the Cygnus Loop is 
extended in almost the entire FOV of the chip (see, Fig.~\ref{fig:Chandra
  image} {\it lower}).  We thus subtract background emission from a
source free region in the S1 chip of the same observation, after we
investigate that the spectral shape and count rate in the energy band
of 3--5\,keV, where the emission from the Cygnus Loop is negligible,
are similar between the S1 and S3 chip.  We should note that the
background-subtraction method was successfully performed by Miyata et
al.\ (2001). We use data in the energy range of 0.5--2.0\,keV for our
spectral analysis. 

\begin{figure}
\includegraphics[angle=0,scale=.6]{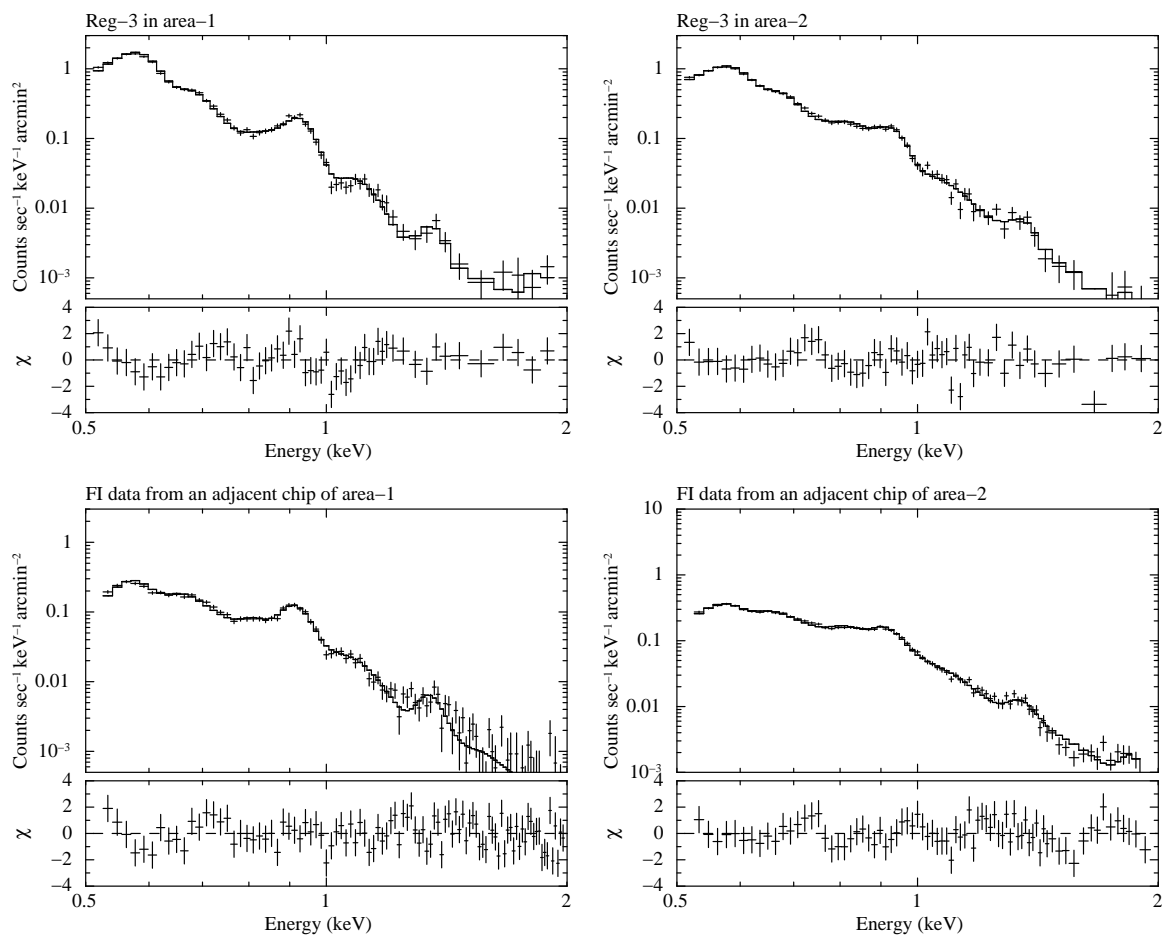}
\caption{Example spectra (crosses) with the best-fit model
  (solid lines).  Lower panel shows the residuals.  {\it Upper left},
  {\it upper right}, {\it lower left}, and {\it lower right} spectra
  are from Reg-3 in Area-1, Reg-3 in Area-2, the northern ellispe in
  Fig.~\ref{fig:HRI image}, and the southern ellispe in
  Fig.~\ref{fig:HRI image}, respectively.} 
\label{fig:ex_spec}
\end{figure}

We apply an absorbed non-equilibrium ionization (NEI) model with a
single temperature (the {\tt wabs}; Morrison \& McCammon (1983) and
the {\tt vpshock} model (NEI version 2.0); e.g., Borkowski et al.\
(2001) in XSPEC v\,12.3.1).  Free parameters are electron temperature,
$kT_\mathrm{e}$; ionization timescale, $\tau$; emission measure, EM
(EM $=\int n_\mathrm{e}n_\mathrm{H} dl$, where $n_\mathrm{e}$ and
$n_\mathrm{H}$ are number densities of electrons and protons,
respectively and $l$ is the plasma depth); abundances of C, N, O, Ne,
Mg, Si, and Fe.  Above, $\tau$ is the electron density times the elapsed
time after shock heating and the {\tt vpshock} model assumes a range of
$\tau$ from zero up to a fitted maximum value.  Based on previous
{\it Suzaku} observations of these regions (Katsuda et al.\ 2008), 
we here fix the values of $N_\mathrm{H}$ and abundance of S to be 0.02
$\times10^{22}$\,cm$^{-2}$ and 0.2 times the solar value, respectively. 
We also set abundance ratios of C/O and N/O to be 2 and 1.4 times the
solar values which are mean values measured in the NE rim of the Cygnus Loop
[calculated from Table~2 in Katsuda et al.\ (2008)], because emission
lines from C and N are not visible in our data. We set
abundances of Ni equal to that of Fe.  Abundances of other elements
included in the {\tt vpshock} model (i.e., Ar and Ca) are fixed to the
solar values (Anders \& Grevesse 1989).  In this way, we fit all the
spectra by the one-component {\tt vpshock} model.  We then find apparent
discrepancies between our data and the best-fit model in the energy
band around 0.73\,keV.  Around this energy band, the dominant 
emission comes from either Fe-L shell lines or K-shell lines from O
{\scshape VII} and O {\scshape VIII}.  However, the poorly 
known atomic physics of Fe-L shell lines and missing O-K shell
lines in the model employed are reported by Warren \& Hughes (2004) and
Yamaguchi et al.\ (2008), respectively.  We thus introduce systematic
error of 5\% on the model in quadrature, following as Warren \& Hughes
(2004) did.  This model give us fairly good fits for all the spectra 
(reduced $\chi^2$ ranges from 0.8 to 1.4).  Figure~\ref{fig:ex_spec}
shows the example spectra in Reg-3 from both area-1 and area-2.  The
spectra are remarkably different; emission lines are stronger 
for the spectrum in area-1 than those for the one in area-2.  The
best-fit parameters and fit statistics for the two example spectra are
summarized in Table~\ref{tab:param}.

\begin{figure}
\includegraphics[angle=0,scale=.7]{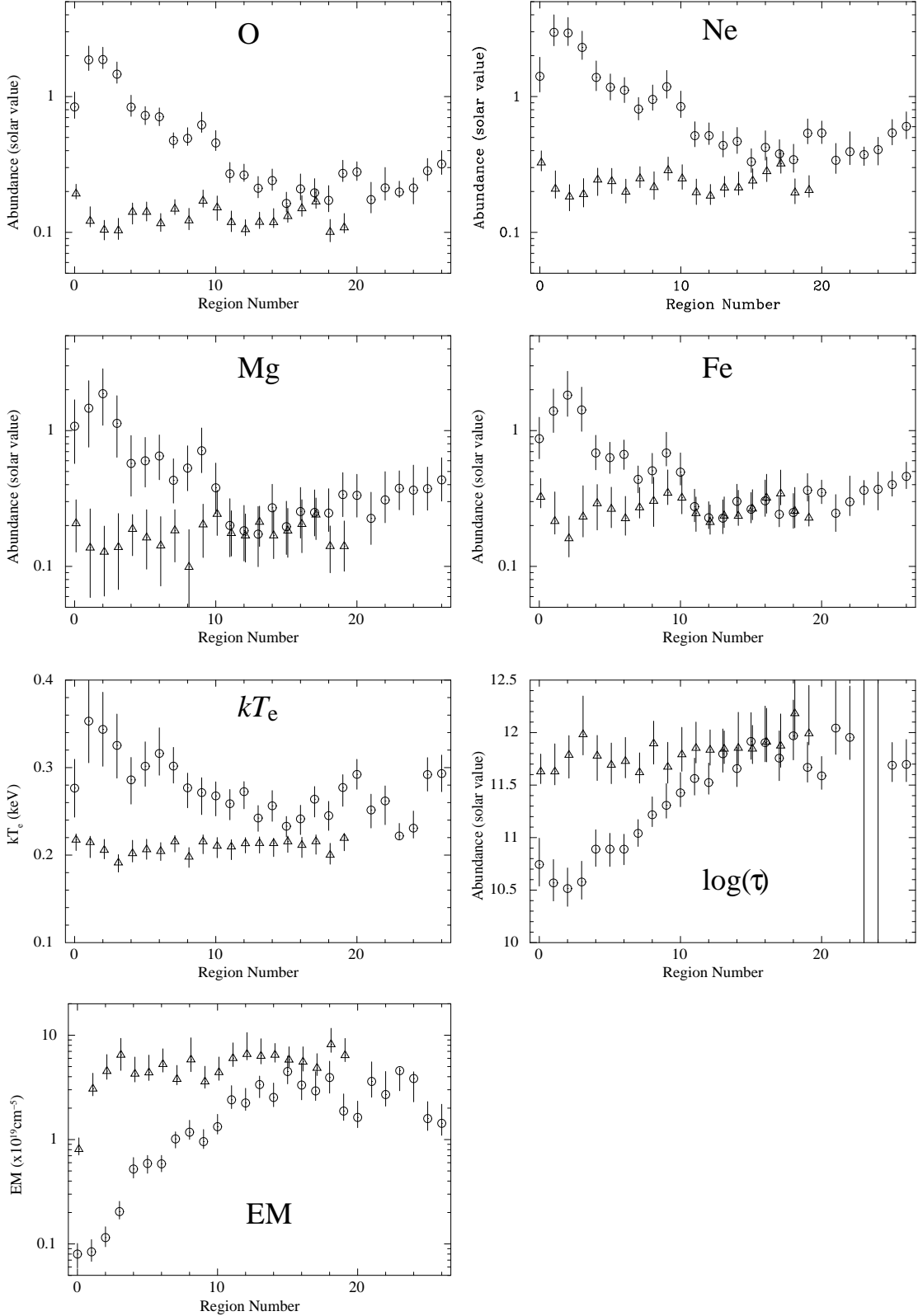}
\caption{Abundance, \kTe, $\tau$, and EM as a
  function of region number.  Open circles represent data points from
  area-1, while open triangles represent those from area-2.
} 
\label{fig:param_dist}
\end{figure}

Figure~\ref{fig:param_dist} shows abundances of O, Ne, Mg, and Fe, 
\kTe, log($\tau$), EM as a function of region number.   
Data points with open circles are from area-1, while those with open
triangles are from area-2.  In area-2, abundances are constant at O$\sim$0.1,
Ne$\sim$0.2, Mg$\sim$0.15, and Fe$\sim$0.2, while in area-1 they 
increase toward the outer region of the Loop.  They start to increase
from around Reg-10 ($\sim$200$^{\prime\prime}$ behind the shock
front) to the outer regions.  Inside Reg-10, the values of 
abundances are the same level as those in area-1, while they become up
to the solar values or even higher than the solar values in the
outermost three regions. The electron temperature is almost constant
at $\sim$0.2\,keV in area-2, while it gradually increases from inner
regions (0.25\,keV) toward outer regions (0.35\,keV) in area-1.  These
values are similar to those of the hot component among the two
components revealed by {\it Suzaku} data (Miyata et al.\ 2007).  
The ionization states for the abundance-enhanced region show NEI conditions
that are consistent with the previous results from {\it Suzaku}.  On
the other hand, we find that the other regions show almost collisional
ionization equilibrium (CIE) conditions.  In contrast
to the electron temperature and abundances, the ionization timescale
as well as the EM gradually decreases toward the outer regions in area-1. 

\begin{figure}
\includegraphics[angle=0,scale=.65]{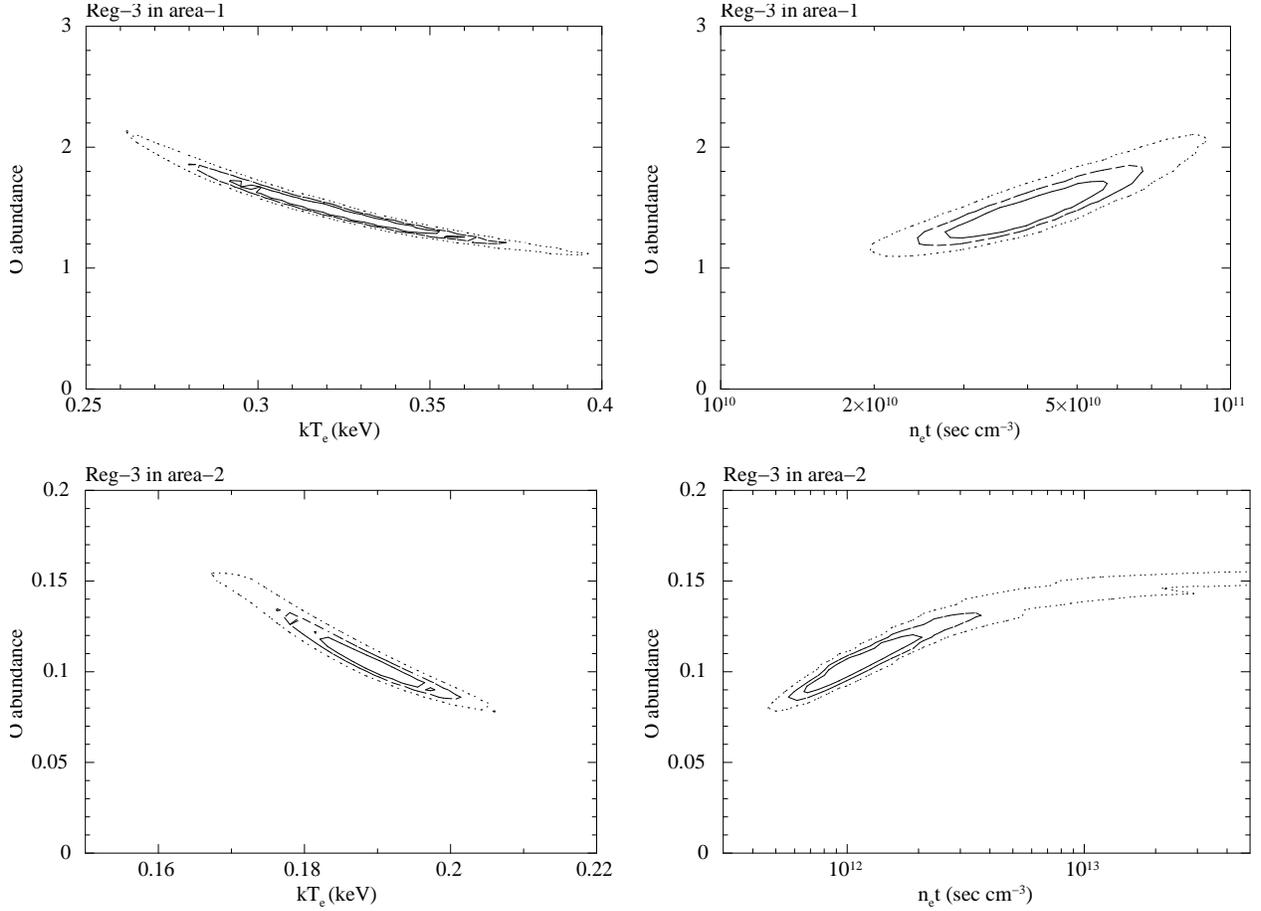}
\caption{Confidence contour maps of O abundance against \kTe~or
  $\tau$.  Upper panels are those for Reg-3 in area-1, while lower
  panels are those for Reg-3 in area-2. Solid, dashed, and dotted
  lines represent 68\%, 90\%, and 99\% confidence levels,
  respectively.} 
\label{fig:conf_cont}
\end{figure}

We can see apparent correlations between abundances and thermodynamic
parameters (i.e., \kTe~and $\tau$).  Therefore, we investigate 
whether the abundance enhancements in the outermost
($\sim$200$^{\prime\prime}$-thickness) region in area-1 are significant or
not, by generating confidence contours between O abundance and the
thermodynamic parameters.  Figure~\ref{fig:conf_cont} shows the
confidence contours of O abundance against \kTe~and $\tau$ for Reg-3
in area-1 (abundance-enhanced region) and Reg-3 in area-2 (the region
with ``normal'' abundances).  We confirm that O abundance for the
abundance-enhanced region is significantly higher than that in the
region with ``normal'' abundances.  We thus conclude that the metal
abundances in the northern outermost region
($\sim$200$^{\prime\prime}$-thickness regions behind the shock front) 
of the NE rim are significantly enhanced relative to those in the rest
of the NE rim.

The analyses above are all concentrated in data taken by the S3 chip.
For the sake of consistency, we analyze data from adjacent FI chips.
We extract spectra from two regions which are located in 
the northernmost chips of ObsID.\ 2821 and 2822, respectively.
These regions are shown in Fig.~\ref{fig:HRI image} as white ellipses. 
Subtracting the background emission from the same chips, we apply the same
{\tt vpshock} model used before.  The two spectra with their best-fit
models are shown in Fig.~\ref{fig:ex_spec}.  The apparent difference
between the two spectra is the intensity of line emission, as seen in the
two example spectra from the S3 chip in each observation (see,
Fig.~\ref{fig:ex_spec} {\it upper left} and {\it upper right}).  The
best-fit parameters and fit statistics are summarized in
Table.~\ref{tab:param}.  Both absolute and relative abundances for the
northern ellipse are almost consistent with those outside Reg-10 in
area-1 (i.e., abundance-enhanced region), whereas those in the
southern ellipse are consistent with those in area-2 (i.e., the region
with ``normal'' abundances).  Therefore, we confirm that the
abundances derived are independent on ACIS chips.

\section{Discussion}

\subsection{Possible Origin of the Plasma in the Abundance-Enhanced Region}

We confirm the enhanced abundances in the northern outermost region in
the NE rim of the Cygnus Loop and reveal that the
abundance-enhanced region is confined in a
$\sim$200$^{\prime\prime}$-thickness region behind the shock front. 
The abundances derived here are higher than those derived by {\it
  Suzaku} by factor of $\sim$2.
The width, $\sim$200$^{\prime\prime}$, of the abundance-enhanced region is
comparable to the half-power diameter, $\sim$120$^{\prime\prime}$, of
the {\it Suzaku} X-ray telescope (Serlemitsos et al.\ 2007) so that
significant amount of emission from the plasma with ``normal''
abundances should have contaminated in the abundance-enhanced region
for the {\it Suzaku} data.  The superior spatial resolution of the 
{\it Chandra} XRT allows us to accurately determine the metal
abundances in the abundance-enhanced region as well as to reveal the
detailed plasma structures there.

Are the abundance enhancements caused by metal-rich ejecta such as
fragments of ejecta in the vicinity of the Vela SNR (i.e., Vela
shrapnels; Aschenbach et al.\ 1995)?  In the Vela shrapnels A and D,
anomalous abundance ratios between heavy elements are observed:
Si/O$\sim$10 times the solar value in the Vela shrapnel A (Katsuda \&
Tsunemi 2005), Ne/Fe$\sim$10 times the solar value in the Vela
shrapnel D (Katsuda \& Tsunemi 2006).  On the contrary, we measure
abundance ratios among metals in the abundance-enhanced region of the
NE rim of the Cygnus Loop to be consistent with the solar values
within a factor of $\sim$2.  Furthermore, the feature that the EM in the
abundance-enhanced region gradually decreases toward the shock front
is in contrast with those observed in the Vela shrapnels A and D.
These facts lead us to consider that the origin of the abundance
enhancements in the NE rim is not likely due to fragments of ejecta.
Therefore, a natural explanation of the abundance enhancements is the
interstellar medium (ISM) origin, since the abundances are roughly
consistent with the solar values.  

We should note that the electron temperature in the abundance-enhanced
region tends to increase toward the shock front.  This feature
resembles those in the Vela shrapnels A and D, rather than that
expected in Sedov phase SNRs where we expect to see temperature 
decrease toward the shock front as seen in the other NE rim of the
Cygnus Loop found in previous observations (e.g., Miyata et al.\ 1994;
Katsuda \& Tsunemi 2008). 

\subsection{Non-Thermal Emission?}

What causes such depleted (typically $\sim$0.2 times the solar value) 
abundances observed in the rest of the NE rim?  One possibility is
resonance scattering which can affect our abundance estimation.
Raymond et al.\ (2003) studied effects of resonance scattering of O
{\scshape VI} photons in the NE rim of the Cygnus Loop, by using FUV
observations.  They concluded that resonance scattering affected O
{\scshape VI} intensities by a factor of 2 level.  More recently,
  Miyata et al.\ (2008) also investigated the resonance-scattering
  effect by their {\it Suzaku} data covering the NE rim of the
  Cygnus Loop.  In their analysis, the abundances derived were all
  depleted to typically 0.23 times the solar values with exception of
  O; O is depleted by an additional factor of two, which is well
  consistent with the results obtained here by {\it Chandra}.  Miyata
  et al.\ (2008) concluded that about a factor of 2 depletion for only
  O abundance can be attributed to resonance scattering, while it is
  not sufficient to account for the abundance depletion observed.
The other possibility is dust 
sputtering (Spitzer \& Jenkins 1976).  However, this mechanism is also
ruled out due to the fact that even Ne is measured to be depleted to
the solar values; Ne is a rare gas so that there is no observational
evidence of Ne depletion. These facts require the other mechanisms for
the origin of the abundance depletion. 

\begin{figure}
\includegraphics[angle=0,scale=.65]{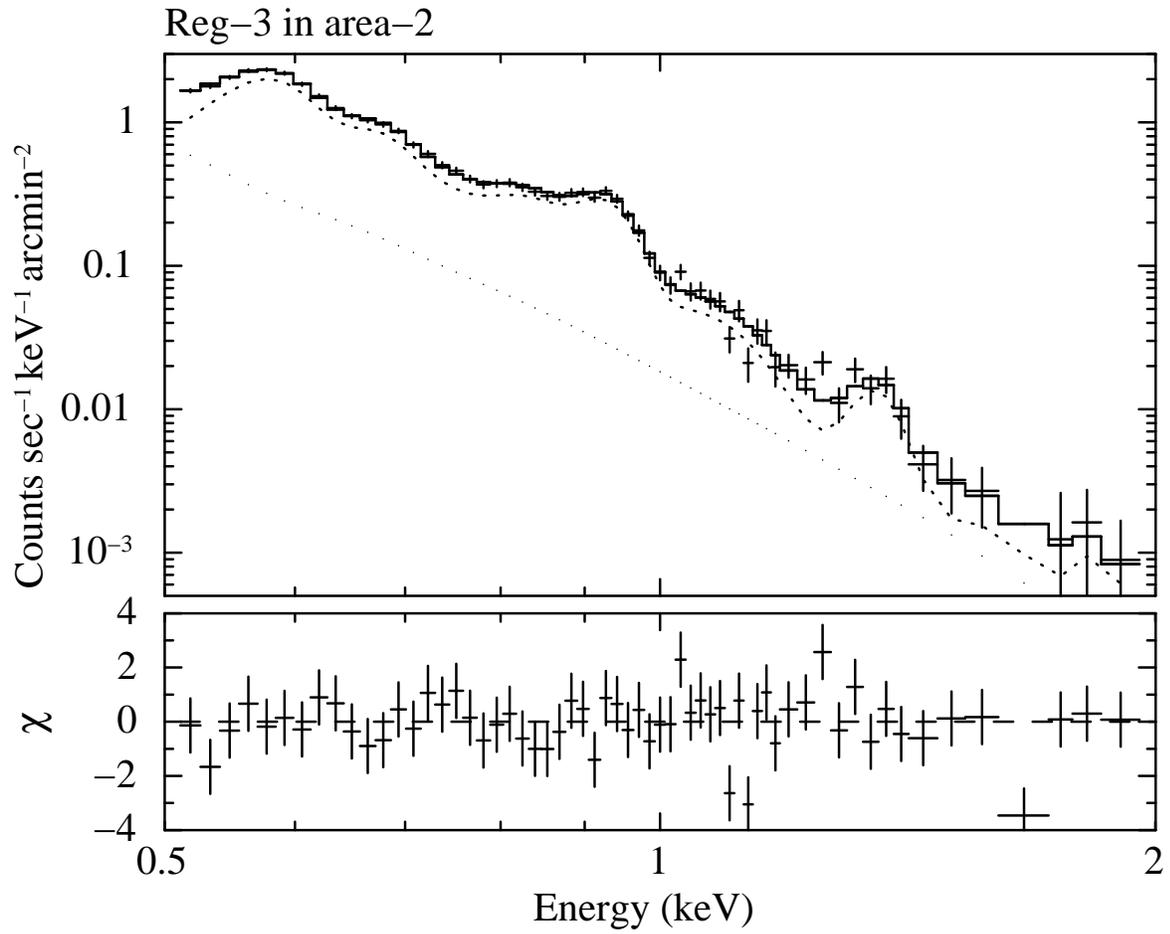}
\caption{Same as Fig.~\ref{fig:ex_spec} {\it Left} but with the
  best-fit {\tt power-law}$+${\tt vpshock} model.  The two
  components are separately shown as dotted lines.
  }  
\label{fig:ex_spec2}
\end{figure}

We here consider effects of non-thermal emission.  If the spectra are
contaminated by non-thermal emission, we should have overestimated
thermal continuum emission in our spectral modeling in which we attempt to
reproduce spectra by only thermal emission.  This will result in
underestimation of metal abundances.  We thus 
investigate whether the spectra require non-thermal emission or not.
First, we introduce a {\tt power-law} model in addition to the {\tt
  vpshock} model for the spectral modeling. In this fitting procedure,
we fix O abundance to the solar value, since absolute abundances
(i.e., relative to H) are not constrained due to difficulty in 
dividing continuum emission into two (i.e., thermal and non-thermal)
components.  Free parameters are photon index and normalization of the
{\tt power-law} component and electron temperature, ionization timescale,
EM, and abundances of Ne, Mg, Si, and Fe(=Ni) in the {\tt vpshock}
component.  Abundances of other elements are fixed to the solar values
(Anders \& Grevesse 1989). In this way, we re-fit all the spectra by
this model.  From statistical point of view alone, this model
significantly improves the fits for some spectra.  As an example, we
show the spectrum from Reg-3 in area-1 with the best-fit model in
Fig.~\ref{fig:ex_spec2}.  The best-fit parameters are summarized in
Table~\ref{tab:param}.  The improvements of this {\tt
  power-law}$+${\tt vpshock} model compared with the previous {\tt
  vpshock} model (see, Fig.~\ref{fig:ex_spec} {\it right}) mainly
comes from the energy band around 0.73\,keV. Since the uncertainty of
the plasma code is reported in this energy range, we should carefully
judge whether the {\tt power-law} component is really required or not.  

In this context, we evaluate multi-wavelength emission.  For this
purpose, instead of the {\tt power-law} model, we employ the {\tt
  srcut} model, which describes synchrotron radiation from an
exponentially cut off power-law distribution of electrons in a
homogeneous magnetic field (Reynolds \& Keohane 1999).  In this model,
we fix radio spectral index to 0.42 (Uyaniker et al.\ 2004), while rolloff
frequency and flux at 1 GHz are left as free parameters.  The other
parameters are treated as the same in the previous {\tt
  power-law}$+${\tt vpshock} model fitting.  The  
best-fit parameters are summarized in Table~\ref{tab:param}.  We find
that the derived best-fit value of the flux at 1 GHz obtained in the
example spectrum (which is extracted from a small portion of the
remnant), 1400 Jy, is about an order of magnitude higher than that
estimated in the entire remnant of about 170 Jy (Uyaniker et al.\
2004).  Therefore, the derived flux of non-thermal emission in the 
example spectrum seems to be unreasonably high.  However, we can not
conclude whether or not non-thermal emission is significant in the NE
rim of the Cygnus Loop, without radio data for the region which exactly
corresponds to our spectral extraction region, as well as a more
sophisticated model for non-thermal emission.

\subsection{Speculation on the Origin of the Abundance Inhomogeneity
  in the Rim of the Cygnus Loop}

It is believed that the Cygnus Loop is a result from a core-collapse
SN (e.g., Miyata et al.\ 1998) which exploded in a pre-existing cavity 
(e.g., McCray et al.\ 1979).  There are a number of evidence that the
blast wave is recently encountered into a rigid wall of the cavity;
the NE rim (Hester et al.\ 1994; Miyata \& Tsunemi 1999), the
southeast rim (Graham et al.\ 1995; Miyata \& Tsunemi 2001), the 
eastern rim (Levenson et al.\ 1996), and the western rim (Levenson et
al.\ 2002).  According to previous X-ray observations, deficient metal
abundances are commonly reported in these regions; metal abundances
are typically $\sim$0.2 times the solar values in the NE rim (e.g.,
Miyata \& Tsunemi 1999; this work), O group abundance is $\sim$0.2
times the solar value in the western rim (Leahy 2004).  In these
regions, the X-ray--emitting plasma should be the shock heated gas
originating from either the cavity wall or 
swept-up matter in the cavity.  Although the reason of the depletion
still remains as an open question, these plasma seem to show depleted 
abundances. 

Meanwhile, there is an apparent break of the cavity wall
that is seen as the south blowout region (see, e.g., figures in
Aschenbach \& Leahy 1999).  The wall of the cavity there is considered
to be so thin (or negligible) that the blast wave already overran the
wall, although Uyaniker et al.\ (2002) proposed another possibility,
i.e., a different SNR interacting with the Cygnus Loop, for the south
blowout region.  The similar structure is expected at the circular
shell in the northwest of the Cygnus Loop along the line of sight
(Tsunemi et al.\ 2007).  Looking at the NE rim of the Loop, we see
somewhat large expansion of the X-ray boundary at 
the abundance-enhanced region (see, Fig.~\ref{fig:HRI image}).
Therefore, we presume that the blast wave here also overran the cavity
wall and now is proceeding into the surrounding ISM with about solar
metallicity, resulting in relatively enhanced abundances there.  In
this context, we predict that abundances show about solar values
around the X-ray boundaries where cavity wall seems to be broken.
This will be checked by future X-ray observations.

\section{Conclusion}

We analyzed archival {\it Chandra} data of the Cygnus Loop NE rim
where abundance inhomogeneities were found by recent
{\it Suzaku} observations (Katsuda et al.\ 2008).  Thanks to the
superior spatial resolving power of {\it Chandra}, we were able to
carry out detailed spatially resolved spectral analyses for the region.
We revealed that the abundance-enhanced region was concentrated
in a $\sim$200$^{\prime\prime}$-thickness region behind the shock
front and confirmed that the values of abundances were consistent with
the solar values by a factor of $\sim$2.  Also, we found that the emission
measure decreased toward the shock front.  These features showed stark
contrast with those seen in the Vela shrapnels, indicating that the
abundance enhancements in the NE rim of the Cygnus Loop were not
likely due to fragments of ejecta.  We suggested that the plasma in
the abundance-enhanced region originated from the ISM, whereas the
plasma in the rest of the NE rim (i.e., abundance-depleted region)
originated from the cavity wall or the gas in the cavity.

\acknowledgments

This work is partly supported by a Grant-in-Aid for Scientific
Research by the Ministry of Education, Culture, Sports, Science and
Technology (16002004).  This study is also carried out as part of 
the 21st Century COE Program, \lq{\it Towards a new basic science:
depth and synthesis}\rq.  S.K.\ is supported by JSPS Research Fellowship 
for Young Scientists.


\begin{deluxetable}{lcc|cccc}
\tabletypesize{\tiny}
\tablecaption{Spectral-fit parameters for example spectra.}
\tablehead{
 & \colhead{Reg-3 in Area-1}& \colhead{Northern Ellipse}&
  &\colhead{Reg-3 in Area-2}& & \colhead{Southern Ellipse}
}
\startdata
Model&{\tt vpshock}&{\tt vpshock}&{\tt vpshock}&{\tt power-law}$+${\tt
  vpshock}&{\tt srcut}$+${\tt vpshock}&{\tt vpshock}\\
&&&&\\
Photon Index$^a$&\dotfill&\dotfill&\dotfill&7.0$^{+0.9}_{-0.5}$&0.42 (fixed)&\dotfill\\
Flux$^b$&\dotfill&\dotfill&\dotfill&3.3$\pm$0.1&1400$^{+5000}_{-1100}$&\dotfill\\
$\nu_{\rm rolloff}$
($\times10^{14}$Hz)&\dotfill&\dotfill&\dotfill&\dotfill&2.9$\pm$0.6&\dotfill\\ 
&&&&\\
$kT_{\rm e}$ (keV)\dotfill & 0.32$\pm0.04$ & 0.32$^{+0.3}_{-0.4}$& 0.19$\pm0.01$ &0.27$\pm$0.05&0.26$^{+0.05}_{-0.06}$&0.23$\pm0.01$\\
log$(\tau /\rm cm^{-3}\,sec)$\dotfill &10.58$^{+0.20}_{-0.16}$&10.90$^{+0.14}_{-0.16}$ &11.8$<$ &11.2$^{+0.4}_{-0.2}$&11.3$^{+0.7}_{-0.2}$&11.5$\pm0.1$\\
O\dotfill& 1.4$^{+0.2}_{-0.3}$&0.8$^{+0.2}_{-0.1}$&0.10$\pm0.02$ &1 (fixed)&1 (fixed)&0.11$\pm0.01$\\
Ne\dotfill& 2.3$^{+0.7}_{-0.4}$ & 1.3$^{+0.5}_{-0.3}$& 0.19$^{+0.06}_{-0.04}$ &1.6$\pm0.2$&1.7$^{+0.4}_{-0.1}$&0.21$\pm0.03$\\
Mg\dotfill & 1.1$^{+0.7}_{-0.5}$ &0.7$^{+0.4}_{-0.2}$ &0.14$^{+0.11}_{-0.07}$ &1.2$\pm0.4$&1.2$\pm0.4$&0.17$^{+0.06}_{-0.05}$\\
Si\dotfill&$<$20&$<$0.3& $<$1 &$<$3&$<$3&0.6$\pm0.3$\\
Fe(=Ni)\dotfill&1.4$^{+0.7}_{-0.4}$&0.75$^{+0.25}_{-0.19}$&0.23$^{+0.16}_{-0.07}$ &1.2$^{+0.3}_{-0.2}$&1.2$^{+0.4}_{-0.2}$&0.21$^{+0.04}_{-0.03}$\\
EM ($\times10^{19}$ cm$^{-5}$)\dotfill& 0.20$^{+0.05}_{-0.03}$&0.37$^{+0.09}_{-0.07}$&6.5$^{+2.9}_{-1.9}$ &0.27$^{+0.15}_{-0.06}$&0.28$^{+0.26}_{-0.08}$&7.0$^{+1.0}_{-0.9}$\\
&&&&\\
$\chi^2$/d.o.f.\dotfill&56/52 &111/92&60/54 &52/52&50/52&62/70\\
\enddata
  \tablecomments{The best-fit parameters for example spectra in
 Fig.~\ref{fig:ex_spec} and Fig.~\ref{fig:ex_spec2}.
 The errors are in the range $\Delta\,\chi^2\,<\,2.7$. 
 Errors for abundances and fluxes in {\tt power-law} and {\tt srcut}
 models are estimated with the EM fixed at the best-fit
 value.  Errors for the EM are estimated with the O abundance fixed at the
 best-fit value.  $^a$Photon index in the {\tt power-law} model is
 determined by the X-ray spectrum, while it is determined in a radio
 spectrum for the {\tt srcut} model.  $^b$The units are
 10$^{-5}$ photons\,cm$^{-2}$\,sec\,$^{-1}$\,keV$^{-1}$ at 1 keV for
 the {\tt power-law} model, and Jy at 1 GHz for the {\tt srcut}
 model. \\ 
  }\\ 
\label{tab:param}
\end{deluxetable}

\end{document}